# Droplet beams for serial crystallography of proteins

U Weierstall[1], R B Doak[1], J C H Spence[1], D Starodub[1], D Shapiro[2], P Kennedy[1], J Warner[1], G G Hembree[1], P Fromme[3] and H Chapman[4]

[1] Department of Physics, Arizona State University, Tempe, AZ 85287, USA
[2] Advanced Light Source, Lawrence Berkeley National Laboratory, 1 Cyclotron Road, Berkeley, CA 94720, USA
[3] Department of Chemistry and Biochemistry, Arizona State University, Tempe, AZ 85287, USA
[4] University of California, Lawrence Livermore National Laboratory, 7000 East Ave., Livermore, CA 94550, USA

email: weier@asu.edu

**Abstract.** Serial diffraction of proteins requires an injection method to deliver protein molecules - preferably uncharged, fully hydrated, spatially oriented, and with high flux - into the crossed beams of an alignment laser and a focused probe beam of electrons or X-rays of typically only a few tens of microns diameter. The aim of this work has been to examine several potential droplet sources as to their suitability for this task. We compare Rayleigh droplet sources, electrospray sources, nebulizers and aerojet-focused droplet sources using time-resolved optical images of the droplet beams. Shrinkage of droplets by evaporation as a means of removing most of the water surrounding the proteins is discussed. Experimental measurements of droplet size, conformation of proteins after passing through a Rayleigh jet, and triboelectric charging are presented and conclusions are drawn about the source configuration for serial diffraction. First experimental X-ray diffraction patterns from Rayleigh droplet beams doped with 100nm gold balls are shown.

PACS: 42.25.Fx, 47.54.Fj, 47.54.De, 47.55.D-, 47.61.-k, 87.15.-v, 87.64.Bx, 07.85.Qe

## 1. Introduction

Electron or X-ray diffraction from a beam of laser aligned molecules has recently been proposed as an approach to structure determination of proteins, particularly those that are difficult to crystallize and are too large for structure determination by NMR [1]. Currently, the structures of $\approx$ 38,000 proteins have been determined by X-ray diffraction or NMR. However structures of several important classes of proteins such as multi-protein complexes or membrane proteins are still rare. For example, less than 120 structures of different membrane proteins have been determined so far. This lack of knowledge of these classes of proteins is in contrast to their biological and medical importance, since 30% of human proteins are membrane proteins and they represent 60% of all drug targets. Similar small numbers of known structures exist for the physiological important multi-protein complexes.

For most large proteins and membrane protein complexes, X-ray crystallography is currently the main method for protein structure determination. However, structure determination by X-ray crystallography depends on the difficult process of protein crystallization. Cryo-electron-diffraction has also been successfully used for structure determination of proteins. Structure determination with NMR avoids the crystallization process; however the applicability of NMR is limited due to the need for isotope labeling, the very high protein concentrations that have to be used, the long data collection time at room temperature as well as limitations concerning the size of the protein. Thus struc-tural biology is in urgent need of a high-throughput method for protein structure determination.

In our method, a beam of hydrated proteins (or possibly microscopic protein crystals) is injected into an interaction region at the intersection of two crossed beams: a diffracting probe beam (continuous electron or X-ray beam) and an alignment beam (continuous polarized IR laser). The interaction region will usually be in a high vacuum environment to allow for evaporative cooling of the droplet beam and to minimize scattering of the diffracting beam by residual gas molecules. In the interaction region the proteins are aligned via torque exerted on the molecule by the electric field **E** of the laser beam, acting on an **E**-field induced dipole moment of the molecule. If N identical aligned molecules are in the interaction region at any instant and if the coherence patch of the probe beam is not appreciably larger than the target molecule, then the probe beam produces a diffraction pattern having N times the intensity of that from a single molecule. If the coherence width is larger than the molecule, interference between two or more molecules in the beam will average out over all the molecules in the exposure. In many X-ray beamlines one can trade coherence for flux and for this method we would operate at the lowest possible coherence and highest possible flux. Since the molecules are constantly replenished (transit time across the interaction region ~200ns), a serial diffraction scheme of this sort allows much longer exposure times than with a static N-molecule crystal. Moreover each molecule receives a dose far below the critical dose that would cause damage at atomic resolution. A transmission diffraction pattern for one orientation builds up at the detector as scattered intensity from successive identical single proteins is recorded. After a useable diffraction pattern has been accumulated, the detector is read out, the alignment laser polarization rotated, and a new diffraction pattern is recorded for a different molecular orientation. In this way a 3-D diffraction dataset can be recorded. From this dataset the charge density of the molecule can be reconstructed with an iterative Gerchberg-Saxton-Fienup algorithm. Alternatively, serial diffraction from a beam of unaligned microcrystallites would yield powder diffraction data, which can also be inverted by iterative algorithms [2]. Hence if the protein of interest can be grown into microcrystals, diffraction data can be obtained even if the crystallites are much too small for conventional X-ray crystallography.

A protein must fold into a specific structure in order to become biologically active. It is the tertiary structure of the biologically active, folded, native protein that is of primary interest (or the quaternary structure for multi-subunit protein complexes) and care must be taken not to alter this configuration ("denature" the protein) during structure determination. In general, it is assumed that water soluble proteins display their native conformation only when hydrated, but the necessary degree of hydration (gram of water per gram of protein) is unclear. A hydration of 0.2 $h$, i.e. 0.2 gram of water per gram of protein, is generally accepted as the threshold value for biological activity, but recent measurements have demonstrated enzyme activity of lyophilized enzymes at hydrations as low as 3% [3, 4]. It is known from cryo-electron microscopy of proteins that a jacket of vitreous ice can also maintain the native protein conformation [5]. Crystalline ice is to be avoided, though, as its formation disrupts protein structure. "Cryo-protectants" can be added to the protein solution to suppress formation of crystalline ice. Electron diffraction from a beam of crystalline ice-balls in vacuum has been reported [6]. Membrane proteins are often solubilized intact in detergent micelles, which faithfully mimic the native lipid bilayer and preserve the tertiary structure. The serial diffraction method outlined above is equally applicable to a solution of micelle-encased membrane proteins as it is to a solution of water-soluble proteins. Indeed, membrane proteins are a subject of primary interest, given the difficulty in probing the structure of these proteins by conventional means.

Transfer of proteins from liquid solution into vacuum without denaturing them is a critical step for the whole method. Not only does this step require that an adequate coating of water or non-crystalline ice be maintained around each individual protein molecule at all times, but also that the injection process not cause thermal or mechanical stress in any degree that would denature the protein [7, 8]. Several droplet beam sources, atomizers, and nebulizers are capable of producing microscopic water droplets containing embedded proteins and injecting these droplets into vacuum without altering the protein conformation. Very few are capable of producing droplets in the quantity and density required.

The number of molecules in the diffraction probe beam at any instant is determined by the size of the interaction region (as set by the focused widths of the probe and alignment beams) and by the density of molecules in this interaction region. To keep exposure times manageable, it is imperative that the flux of analyte molecules into the interaction region be as large as possible, corresponding to a high droplet beam brightness. Accordingly the protein-doped droplets emitted from the molecular injection source should be collimated (preferably a single file train) and unidirectional.

At optical frequencies (e.g. the near-IR laser frequency used for molecular alignment), experimental measurements of refractive index have shown that biomolecules can be described as homogeneous bodies with dielectric constant of 2 – 2.3 [9, 10] This being the case, the induced dipole moment of a protein is mostly due to the shape anisotropy of the molecule. At low frequency, polarization is due to the re-orientation of polar molecules. The interior of proteins can be treated as homogeneous with a low dielectric constant, but a large polarization arises on the highly inhomogeneous surface due to the side-chain re-orientation [11]. Moreover in a DC field, permanent dipole moments (which cancel out in AC fields) can produce large alignment forces. Therefore strong electrostatic [12] and magnetic fields [13] may be used alone or in combination with CW laser fields for molecular alignment. Flow alignment due to shear is also possible, as used in studies of electric birefringence, and this may well occur in the Poiseuille-like flow expected in the droplet beam nozzle.

The electrical polarizabilities of water (or ice) and proteins are similar at optical frequencies. Embedding the proteins in spherical water droplets therefore reduces the shape anisotropy of the whole and compromises the efficiency of laser alignment. Accordingly, an ideal injection system would remove almost all water from the protein-doped water droplets, leaving a hydration jacket only a few monolayers thick. This jacket will quickly freeze to ice as the droplets enter the vacuum chamber, whereafter sublimation proceeds only very slowly. The water vapor load on the vacuum system is therefore easily manageable.

To achieve adequate molecular alignment with either electric or magnetic fields, the thermal population of higher rotational states has to be minimized, i.e. the molecules have to be cooled to cryogenic temperatures. This can be achieved, e.g., by free-jet expansion into vacuum of target molecules mixed into helium gas or by evaporative cooling augmented by quench cooling in co-flowing 4K helium gas.

In summary, serial diffraction requires an injection method to deliver protein molecules or microcrystals - preferably uncharged, fully hydrated (in detergent micelles if membrane proteins), spatially oriented, and with high flux - into a focused probe beam of electrons or x-rays of typically only a few tens of microns diameter. The aim of this work has been to examine several potential droplet sources as to their suitability for this task. In doing so we have discovered new aspects of these droplet beams. This paper presents the experimental measurements, draws conclusions about serial diffraction source configuration, and discusses the interesting droplet beam phenomena that have emerged along the way.

**2. Imaging of Droplets**

In order to image the droplets produced by various droplet sources, a simple and inexpensive nanosecond flash illumination system was developed, which can work in both stroboscopic and single exposure mode with exposure times ≤10ns. The droplets produced by a periodically triggered Rayleigh droplet beam (see below for details) can be imaged with a light optical microscope and stroboscopic illumination, where the illumination frequency is synchronized with the breakup frequency of the droplet source. Our experimental setup is shown schematically in Figure 1.

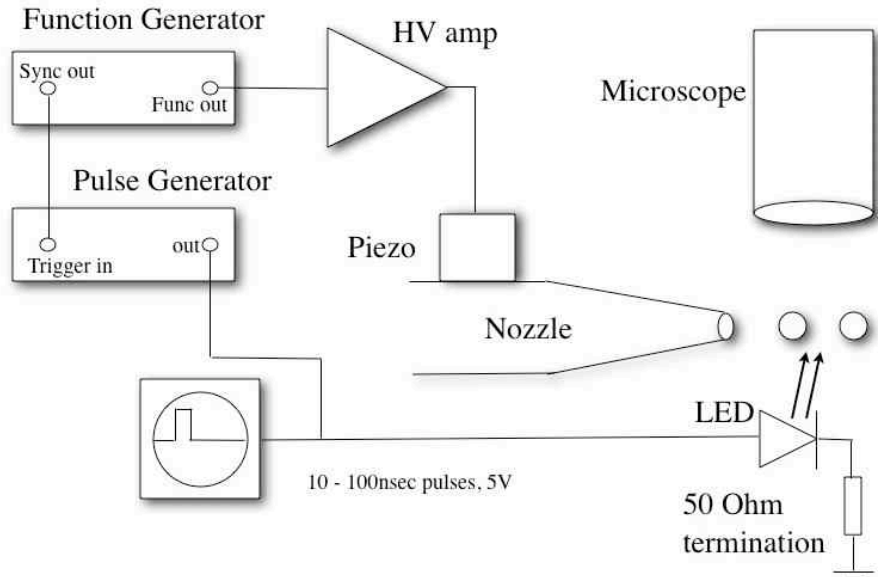

Figure 1: Experimental setup for stroboscopic and single shot droplet imaging using a microscope (Nikon measuring microscope, magnification x200) and pulsed illumination.

We have used fused silica and glass capillary tips from New Objective as nozzles for the Rayleigh jets, as well as fused silica nozzles "pulled" in our lab using a Sutter P-2000 puller [14]. Droplet beams have been formed to date of HPLC water, methanol, aqueous solutions with 100nm gold balls, aqueous solutions with e-coli bacteria and protein solutions. The liquid of interest is pressurized in a PEEK or stainless steel reservoir to ≤400 psi. This reservoir is connected to the nozzle via 1/8" PEEK tubing and the liquid passes through a 0.5 μm PEEK filter before reaching the nozzle. A multilayer piezo-actuator (PL022.31 from Physik Instrumente) with a resonance frequency of about 300 kHz is attached with a clip to the side of the glass nozzle. The actuator generates the periodic disturbance necessary to trigger the Rayleigh breakup of the liquid jet into mono-dispersed droplets.

The piezo-actuator is driven with a sinusoidal voltage from a function generator (Wavetek 2 MHz model 20), amplified by a custom, high frequency, high voltage amplifier (built around an APEX PA85, which has a power response up to 2MHz). The function generator also triggers a pulse generator (Tektronix PG501A), which drives a fast LED (IF-E99 from Industrial Fiber Optics, 3 ns rise time) with 10 - 100ns wide, 5V pulses. A series resistor (not shown) limits the current through the LED to the maximum current allowed. A resistor in parallel with the current limiting resistor and the LED produces a 50 Ohm termination of the transmission line to prevent degradation of fast pulses due to reflections. Since the LED voltage is synchronized with the breakup frequency of the droplets, the LED light acts as stroboscopic illumination, providing a stable image of the droplets over even hundreds of LED flashes [15, 16]. The droplets were observed with a Nikon Measuring Microscope MM20 with bright field illumination and x200 magnification. Images are recorded with a Princeton 1024x1024 liquid nitrogen cooled CCD camera.

Fast imaging of nonperiodic events (e.g. an untriggered Rayleigh droplet breakup) is possible by replacing the LED with a laser diode (Sharp GH06550B2B1, 50mW, $\lambda = 656 nm$) and using the function generator in single pulse mode [17]. Due to the higher power requirements of the laser diode, a high speed pulse amplifier was built [18]. This amplifier also enabled overdriving of the laser diode to gain more light output, which is possible for short pulse durations without damaging the diode. The higher power of the laser diode together with the low noise of the liquid nitrogen cooled CCD camera makes single shot imaging with 20-100ns exposure times possible. The laser illumination produces a strong coherent scattering background (speckle), which makes background subtraction mandatory. This is done by recording and then subtracting a second image, in which the droplet beam has been moved out of the field of view.

**3. Rayleigh droplet beams**
If a liquid is pressurized and forced through a small orifice into vacuum or stagnant air, the liquid emerges as a single, contiguous, cylindrical liquid beam. The overall surface free energy of the cylindrical beam is higher that that of an equivalent flow of spherical droplets. Consequently, the cylindrical jet breaks up spontaneously at a certain critical distance downstream of the source aperture, forming a line of spherical droplets [19, 20]. This instability was first analyzed in the late 1800's, notably by Rayleigh [21], whose name it now bears. The instability can be triggered by perturbing the source conditions (temperature, pressure, etc.) at a frequency near the intrinsic instability frequency. Laminar liquid flow through an aperture into vacuum, triggered in this fashion, yields a unidirectional, monodisperse beam of spherical droplets that is locked in phase to an external signal. Rayleigh breakup is labeled either "spontaneous" or "triggered," respectively, depending on whether it is induced by random acoustic fluctuations or by an intentionally applied excitation. In either case, the nozzle diameter D sets the basic droplet size. Spontaneous breakup yields a narrow but not monodisperse size distribution of droplets, centered on 1.90D droplet diameter and 4.55D spacing on centers. Periodic triggering delivers, as shown in Figure 4, a stream of identical droplets that is highly periodic, monodisperse, monoenergetic, and phase-locked to the excitation signal (Within the ink-jet printer community, this form of droplet beam is generally referred to as a "continuous ink jet" (see, e.g. [22])). By varying the trigger frequency, the droplet diameter of a triggered Rayleigh droplet beam can be fine-tuned from below 2.0D to over 2.5D, with corresponding spacing of 5.3D to 10.4D on centers, respectively. The requisite trigger frequency is $f = v/\xi$, where v is the beam velocity and $\xi$ the breakup segmentation length. Beam velocities from well under 10 m/s to over 100 m/s are readily achievable.

Vacuum experimentation on filamentary liquid jets dates at least to Siegbahn in 1973 [23]. Research on and with such jets has been conducted by many groups, including Keller [24], Faubel [25-30], Mafuné and Kondow [31, 32], Buntine [33], and Saykally [34]. Rayleigh droplet beams have been employed as targets for nuclear physics [35] and x-ray generation [36-41] and as possible fuel sources for fusion reactors [42, 43]. All of these applications require a moderately good vacuum, typically $10^{-4} – 10^{-5}$ Torr or less. All of the above liquid jet experiments have employed nozzles of 5μm diameter or larger, generally much larger.

We use commercially available electrospray tips, which are available in diameters down to 1μm [44] as nozzles for generation of Rayleigh jets. We also pull nozzles ourselves from fused capillary tubing using a Sutter P-2000 puller [14]. The nozzle is connected with PEEK tubing and PEEK fittings (HPCL components [45]) to a PEEK liquid reservoir (or alternatively with Swagelok fittings to a stainless steel reservoir). Immediately upstream of the nozzle is a 0.5μm PEEK inline filter (Upchurch). A photograph of one version of our droplet beam source is shown in Figure 2 (here with stainless steel fittings). Our experiments have shown that nozzle clogging is the most serious problem with Rayleigh jets from nozzles of diameter smaller than 8 μm, regardless of filter size. We clean the whole assembly (in line PEEK filter, liquid reservoir, PEEK tubing, fittings and ferrules) in Micro 90, filtered methanol and filtered HPLC water.

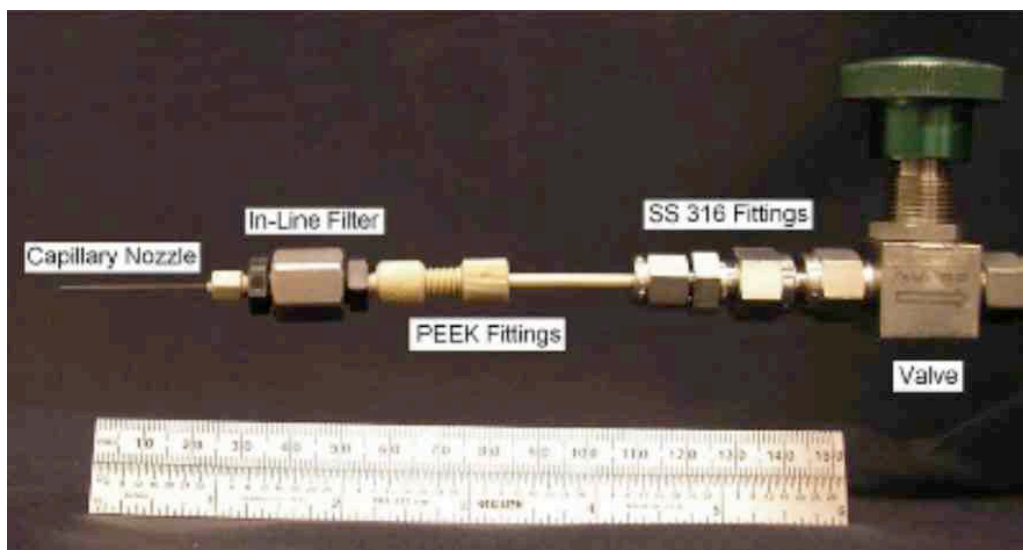

Figure 2: Source assembly. Six inch ruler shown for scale.

Reliable Rayleigh jets can be obtained using online fused silica Picotips® with 50μm ID at the distal end and 8μm diameter at the tip end and with a 0.5μm PEEK filter in line with the nozzle. Using the same nozzle with a 100nm gold ball solution (Ted Pella) also results in a reliable jet. Using the nozzle with detergent buffer used for the membrane protein photosystem I, a protein complex which catalyzes the light-induced charge separation in photosynthesis [46], resulted in a very low success rate because the nozzles tended to clog after a few minutes. It is at the moment unclear if this clogging results from clumping of microparticles in the buffer or from the physical action of the detergent on remaining dirt particles located in the volume between the filter and the nozzle. The detergent buffer could remove particles from surfaces, which would stay attached to the surface if other liquids were used. These particles would then be flushed to the nozzle end and would clog the nozzle. Using an aqueous solution containing a high concentration of 100nm latex spheres [47] resulted in clogging due to particle agglomeration, since these particles are hydrophobic by nature and will always tend to agglomerate, especially at high concentration. In this case the jet became weaker and weaker in a few minutes (observed in an optical microscope at mag 200x) and showed a slow buildup of particles at the tip end. Using smaller nozzle sizes, e.g. 4μm offline glass Picotips (New Objective) resulted generally in lower success rate and higher clogging rate. Again pure water showed the highest success rate, followed by gold ball solutions and methanol. The larger nozzles could occasionally be cleaned (de-clogged) by reverse pressurizing with water at 400psi from the tip end. In order to determine the nature of the observed clogs, X-ray analysis can be used. Figure 3 shows a Scanning Electron Microscope (SEM) image taken "head-on" of a clogged nozzle. The nozzle was a 4μm diameter commercial borosilicate glass nozzle that ran for 4-5 hours with a water jet before clogging. The clog formed somewhat upstream of the nozzle exit. By breaking off the tip of the nozzle, the clog could be exposed for SEM imaging and Energy Dispersive X-ray (EDX) spectroscopy. A 0.5μm pore inline PEEK filter just upstream of the nozzle should remove particles of this size, indicating that the clog material originated within the nozzle tube itself. An EDX spectrum taken on the clog (small circle in Figure 3, left image) shows only Si (light elements, oxygen and below, would not appear in this EDX spectrum). By moving the electron probe to the glass wall, a completely different spectrum was obtained. EDX on the glass wall of the nozzle is dominated by an Al peak with only a much smaller Si peak present. The clog material is therefore chemically different from the wall material and does not consist of a PEEK particle either, since PEEK contains only C, H, and O.

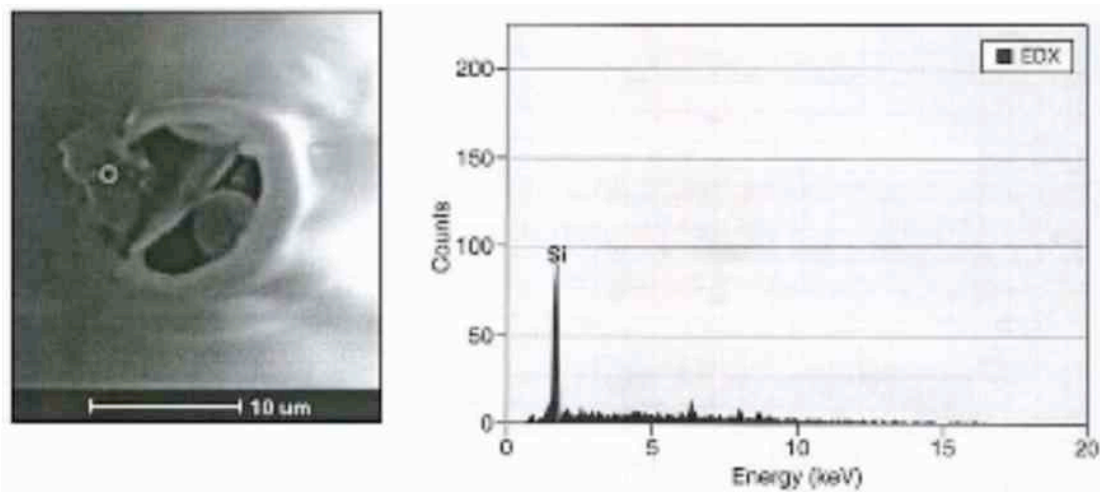

Figure 3: SEM image (left) and EDS spectrum (right) of silicate blockage in a borosilicate glass nozzle.

Figure 4 shows a stroboscopic image of a Rayleigh jet from a 4μm diameter glass nozzle where the droplet breakup is triggered at 1.2 MHz. The monodispersivity of the droplets has also been confirmed by laser diffraction from a triggered Rayleigh beam as shown in Figure 5 (left). Here, the green laser is focused onto the droplet beam resulting in a Airy's disk diffraction pattern from the droplets which is equivalent to the pattern, that would be obtained from one droplet (the high speed movement only appears as a phase factor in the diffracted wave and cancels out when recording the intensity on a screen). On the right, the laser was slightly defocused, which results in two droplets being illuminated at any instant, producing Young's fringes from interference between the waves scattered from two droplets.

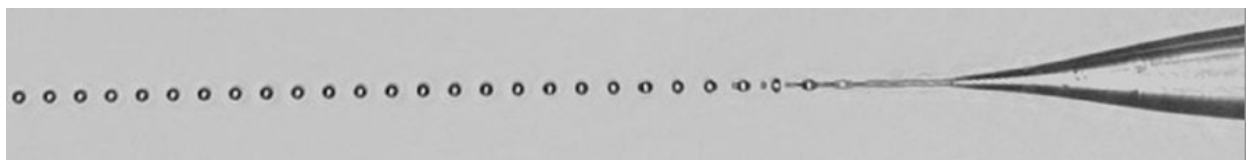

Figure 4: Triggered Rayleigh water jet in air. The droplet diameter is 8 μm (4μm nozzle). A piezo is vibrating the nozzle at 1.2 MHz, triggering the breakup into monodispersed droplets. Water pressure: 200psi. The image was obtained with 100ns stroboscopic flashes (LED) synchronized to the piezo frequency. (Mag: 200)

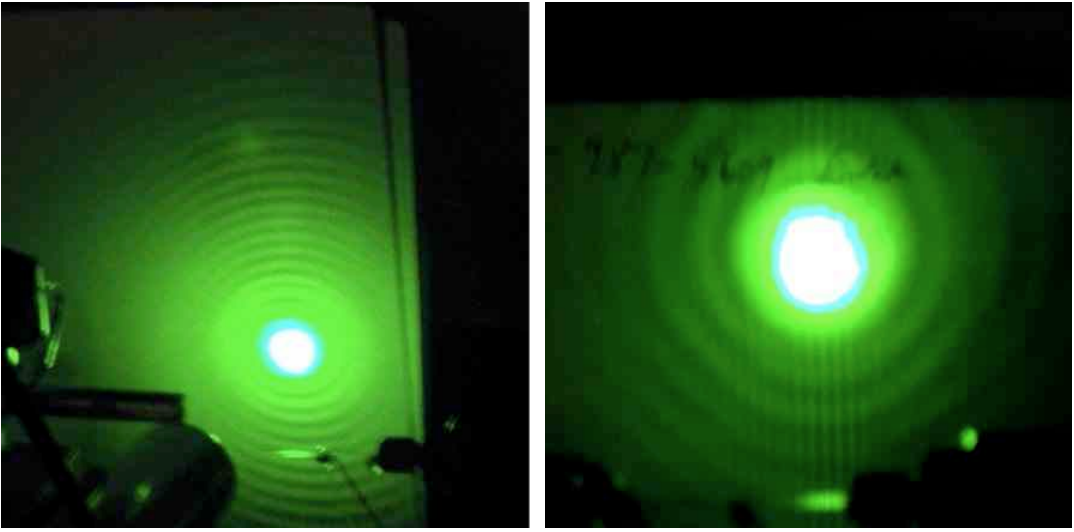

Figure 5: Laser diffraction pattern obtained from a stream of triggered single-file monodispersed droplets of water (8μm diameter) showing an Airy's disk pattern. On the right: Interference fringes between neighboring droplets produce additional Young's fringes.

As shown above, periodic, monodispersed, linear droplet beams can be readily produced, but nozzle clogging problems limit the minimum size of droplets. Droplets of 16μm diameter are easily produced, droplets of 8μm diameter are reasonably straightforward (50% success rate), 4 and 2 μm diameter droplets are very difficult (although the variance in the success rate is large), and one micrometer or sub-micrometer droplets are not possible at present. Since our aim is to remove most of the water surrounding the molecules before injecting them into vacuum, it is desirable to start with the smallest initial droplet size possible.

## 4. Droplet evaporation

Reduction of the cooled droplet size due to evaporation in vacuum is a small effect. This is due to evaporative cooling, since the evaporative flux from a liquid scales with surface area and vapor pressure, and the latter decreases dramatically with temperature. As a result, evaporative cooling of a liquid jet in vacuum limits the overall evaporative flux from the beam, provided the initial surface area of the jet is sufficiently small [27, 30] (i.e. for nozzle sizes of a few micrometers for water).

The rate of evaporative cooling can be estimated assuming: (1) free-molecular flow away from the droplet surface, (2) uniform temperature within the droplet, (3) liquid loss from the droplets by evaporation from the surface at the rate determined by equilibrium vapor pressure at the momentary droplet temperature, and (4) evaporative cooling of the droplet via the latent heat associated with this evaporative flux [30]. The resulting equations for droplet temperature $T$ and radius $r$ as a function of the distance $z$ from a nozzle tip along the jet direction are

$$\frac{dT(z)}{dz} = -\frac{3\Lambda}{c_p r(z)} \cdot \frac{dr(z)}{dz} \quad (1)$$

$$\frac{dr(z)}{dz} = -\frac{P[T(z)]}{v\rho} \cdot \sqrt{\frac{m}{2\pi k_B T(z)}} \quad (2)$$

where $\Lambda$ is the heat of vaporization, $c_p$ is the specific heat capacity, $m$ is the mass of a water molecule, $\rho$ is density of water, $v$ is the droplet velocity, and $k_B$ is the Boltzmann constant. Equation (1) describes the heat transfer balance for a spherical droplet, and Equation (2) relates the rate of the droplet size change to the vapor flux from the droplet surface. Neglecting the effect of droplet curvature, the temperature dependence of saturation vapor pressure over liquid water at temperatures below 273 K is described by the empirical Goff-Gratch equation [48]. The results of calculations for water droplets of several initial diameters, traveling at 37 m/s through perfect vacuum, are presented in Figure 6 and Figure 7. Even a large droplet of 20μm diameter has super-cooled to almost 230 K only 10 cm downstream of its injection point and, as a result, has largely stopped evaporating and shrinking. Experimental measurements verify these predictions [30, 34], indicating cooling rates for water droplets of up to $10^6$ K/s and super cooling to droplet temperatures as low as 207 K. Due to the extremely fast evaporative cooling immediately after droplet injection into vacuum, the dependence of the droplet temperature on the liquid temperature inside the nozzle becomes negligible after the droplet travels only a few centimeters.

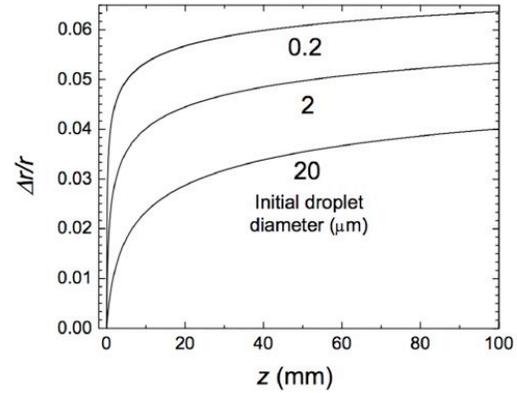

Figure 7: Relative change of droplet size by evaporation in vacuum for the same droplet sizes as in Fig. 6. Due to rapid super-cooling of the droplets (see Figure 6), the droplet size changes only by a few percent.

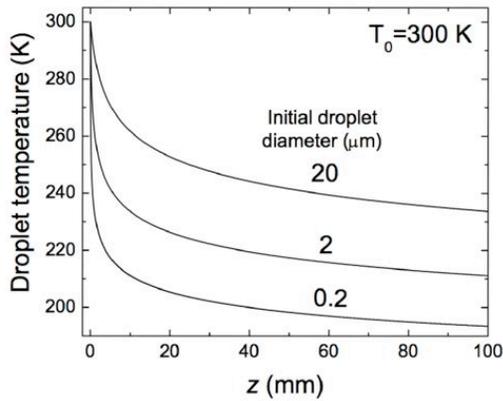

Figure 6: Evaporative cooling of water droplets with different diameters (20, 2, and 0.2 μm), traveling at 37 m/s in vacuum.

Given this self-limiting evaporation, the overall change in size of a droplet traveling through vacuum is relatively minor, as shown in Figure 7. A 2μm water droplet supercooled by 50 K shrinks only about 5% in diameter. To shrink the water jacket of protein-doped water droplets, it will therefore be necessary that most of the water be removed before injection into vacuum or that the initial droplet size is comparable to that of the embedded molecule or microcrystal.

For a Rayleigh droplet jet injected directly into vacuum (as in our first X-ray experiments shown below), the evaporative gas load from a complete droplet beam is easily predicted by summing the evaporative flux along a train of droplets. Within our simple modeling and ignoring re-capture of gas molecules by neighboring droplets, the dependence of the resulting pressure in the vacuum chamber on nozzle diameter is shown in Figure 8. The droplet size and spacing are taken to be the most probable Rayleigh breakup values. This plot shows that the evaporative gas load and accordingly the pressure scale as approximately the square of the nozzle diameter. For a nozzle diameter of 1 micrometer a very modest 600 l/s vacuum pumping speed is predicted to deliver about $10^{-5}$ Torr water pressure. This corresponds to a mean-free-path length of 0.6 m for water molecules, or well into the rarefied gas regime for any normal-sized vacuum chamber. Because most of the evaporation occurs in the vicinity of the nozzle tip, the pressure in the vacuum chamber can be easily reduced further by using differential pumping stages. These calculations are in accordance with actual experimental observations [30, 31, 34] and our own observations with a nozzle system in a soft-X-ray diffraction camera (see below).

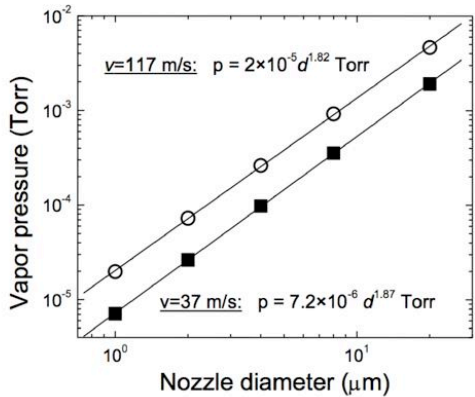

Figure 8: Calculated vacuum chamber pressure at 600 l/s pumping speed for water droplet beams as a function of nozzle diameter *d*, shown at two droplet velocities.

## 5. Triboelectric charging

Triboelectric charging of a fluid flowing through a channel is a phenomenon well known to the chemical industry [49]. Also called "electrokinetic" charging, it results in general when both a velocity gradient and spatially dissimilar distributions of positive and negative charge carriers are present in a liquid. Charging can even occur in pure water, where the charge carriers are simply dissociated H+ and OH- ions. A non-uniform charge distribution always arises spontaneously in a polar fluid near a solid wall due to formation of an electrochemical bilayer at the fluid-wall interface [33]. It is exactly in this near-wall region where the velocity gradient is largest, leading to very effective triboelectric charging. The charging increases with fluid flow velocity [49]. That such charging persists to microscale fluid flows is known from measurements by Faubel with water jets [28] and Buntine with methanol jets [33]. Indeed, being an interfacial phenomenon and therefore dependent on the surface-to-volume ratio of the fluid stream, it is expected to be relatively more important the smaller the diameter of the stream. Charging and spatial dispersion are immediately evident when a Rayleigh droplet beam of 4μm diameter water or methanol droplets is passed through ambient air (room temperature and 1 atm). With heat supplied by the ambient air, the droplets shrink by evaporation without cooling. If charged, they eventually reach the Rayleigh electrostatic limit [19] and undergo "Coulomb explosion," disintegrating into a multitude of smaller charged droplets. Slowing of the droplets via drag forces further enhances Coulombic dispersal of the beam. This dispersal is very much evident in the photograph of Figure 9. It bears similarities to the electrospray disintegration process [50] but is purely triboelectric, since there is no electric potential applied to the nozzle as in electrospray. A current measurement between the source assembly and a plate electrode in the liquid jet shows a streaming current of a few tens of nanoamperes. This current increases with nozzle pressure, namely with increasing flow rate of liquid through the nozzle. The current depends on ph, according to which the sign of the current may be reversed. Electrostatic deflection measurements verify that the droplets of Figure 9 are charged (16 μm droplets charged to $10^6$e), even in the complete absence of any electric potential applied to the nozzle.

Although of great potential interest for mass spectrometry of proteins, triboelectric droplet charging and Coulombic dispersal is detrimental to protein diffraction studies since it may destroy the droplet beam collimation and the charging may change the protein conformation. Triboelectric charging can always be minimized by operating at sufficiently low flow velocities, or by using a ring electrode downstream of the nozzle as used in continuous ink jet printers to charge the droplets [51]. When a voltage is applied to the ring electrode with respect to the jet, polarization charges are induced on the jet. When the end of the polarized liquid column detaches to form a droplet, a net charge is carried downstream. Consequently, the voltage on the ring electrode and the value of the capacitance formed between the jet and the electrode determines the current carried by the droplet beam. By controlling this voltage via a feedback loop based on the droplet beam current, the beam current can be locked in to any desired value, including zero current. Control of the droplet charge is easily observed experimentally since uncharged droplets have a much longer collimated beam range in air than do charged ones.

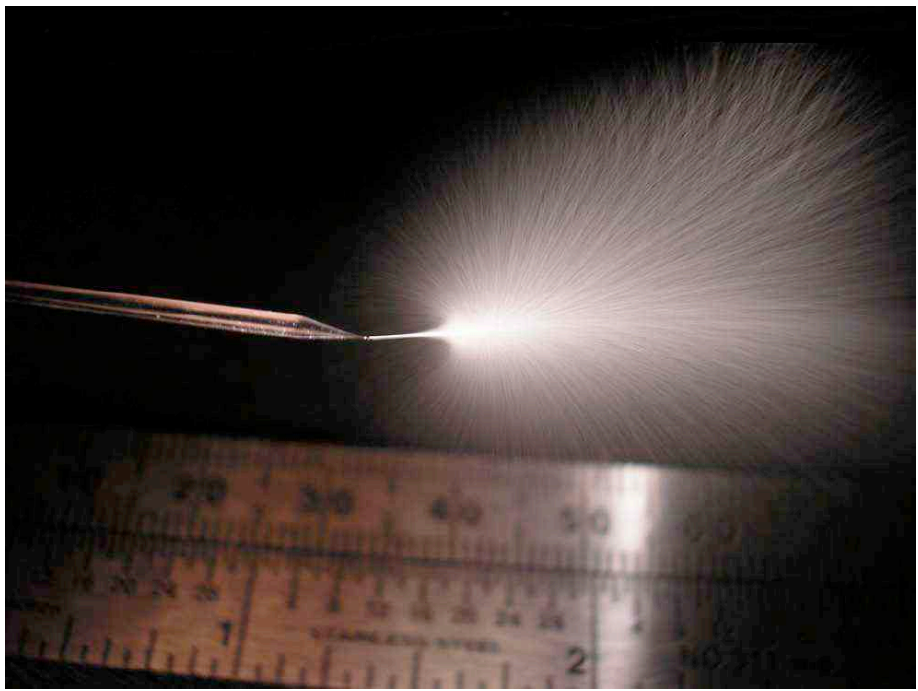

Figure 9: Extended exposure (ca. 1/6 s) of an untriggered Rayleigh droplet beam of methanol issuing (slightly off-axis) from a 2μm diameter glass nozzle into room temperature air. It abruptly disperses about 1 cm downstream of the nozzle due to triboelectric charging resulting in Coulomb explosion and Coulomb repulsion. There is no electric potential applied to the nozzle. Similar effects are seen with water.

**6. Coaxial Gas flow source**

While pursuing solutions to the clogging problem for small diameter nozzles, we also considered the possible use of larger nozzle diameters (~8 μm) to produce large (~16 μm) protein-containing water droplets that must then be shrunk down via evaporation in an ambient gas before they are injected into vacuum. As we have seen above, this evaporation and shrinkage must take place in an ambient gas, since water droplets injected directly into vacuum super cool so rapidly that they shed very little mass before they either cease to evaporate significantly or actually freeze.

The challenge is therefore to pass the droplet beam through a high pressure, high temperature ambient gas without disrupting the single-file, unidirectional form of the water droplet beam. In particular, our measurements on water droplets injected into ambient air verify (as anticipated theoretically) that micrometer-sized droplets rapidly decelerate under these conditions. As the droplet speed approaches zero, the motion of the droplets is increasingly influenced by diffusion and convection, and the desired single-file, unidirectional and monodisperse beam form is lost. The loss of monodispersivity due to air drag can be seen in Figure 10, which shows a single shot image (100ns exposure time) of a triggered Rayleigh droplet beam in air, at several different positions in the beam. Just downstream of the nozzle, the droplets are perfectly monodisperse and equidistant, but due to deceleration in air and wake effects, the droplets lose those attributes.

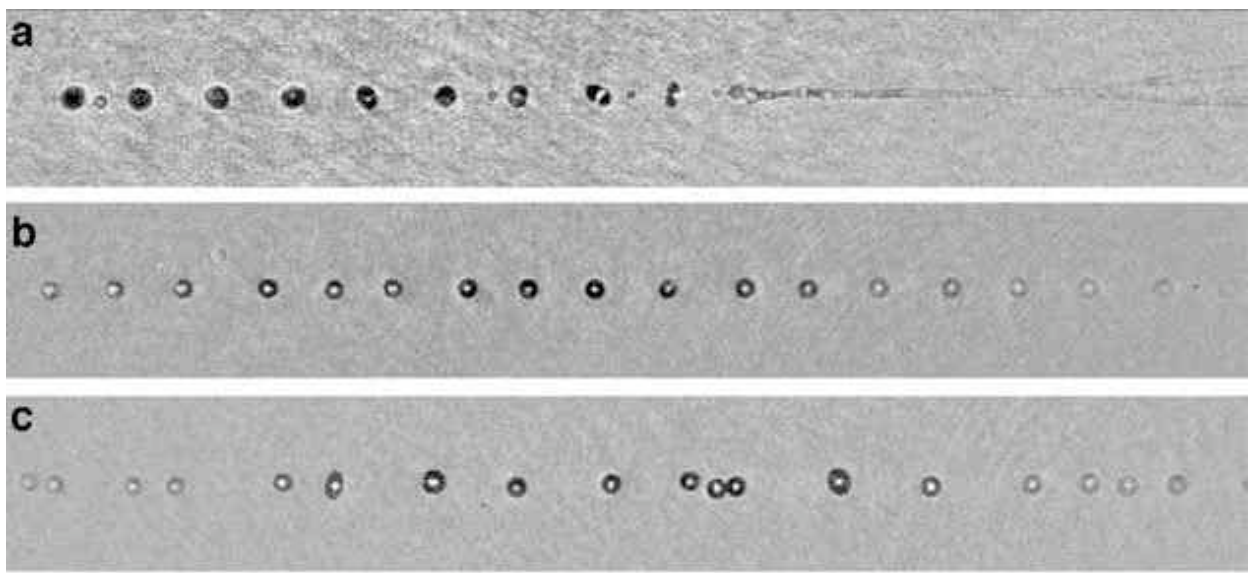

Figure 10: Loss of monodispersivity of a triggered Rayleigh jet in ambient air. Nozzle diameter: 8 μm. Trigger frequency: 1.6MHz, single shot images with 100ns exposure time (a) droplet breakup at the nozzle. (b) 0.5 cm downstream of nozzle. (c) 1 cm downstream of the nozzle. Due to wake effects and air drag some droplets have coalesced, and a slight angular spread is already observable.

If the droplets are charged, either intentionally for electrically forced extraction from the nozzle (electrospray, see below) or by intrinsic electrokinetic (triboelectric) charging of the liquid as it passes through the nozzle, this dispersal is accelerated due to electrostatic repulsion (see e.g. Figure 15). In particular, a line of charges of the same sign is unstable with respect to an undulatory perturbation.

If the ambient gas were moving with the droplets, and if the gas motion were uniform, laminar, and unidirectional, it is possible that the desired linear beam form could be preserved long enough for the droplets (at least neutral droplets) to shrink and be injected into vacuum while still maintaining the desired linear beam.

On the basis of symmetry, one can argue that the most beneficial geometry for the ambient gas flow is one that has the same cylindrical symmetry as the water jet itself, namely a coaxial flow. Flows of one liquid stream within a coaxially flowing sheath of a second liquid are fairly common within the fluid dynamics community. If the velocities of the two liquids are not too disparate, the flow remains laminar. The two liquids then can flow downstream with a distinct inner jet of one fluid surrounded by a coaxial sheath of the other. The radius of the inner jet can be manipulated by adjusting the overall radius of the flow tube and the relative speed of the coaxial sheath. If the flow accelerates, the inner jet must narrow in diameter (a simple consequence of conservation of mass), which may eventually be useful for generating small jet diameters.

With the above in mind, we have initiated work to investigate the effects of a coaxial gas flow co-moving with the droplet beam. The size distribution, directionality, and speed of the droplets will be measured as a function of the speed and species of the ambient gas. A drawing and a photo of the source for our coaxially shrouded water droplet beam are shown in Figure 11. The analyte liquid enters at the upper rear of the main body of Figure 11, and supplies a capillary nozzle of exactly the same variety as we currently employ.

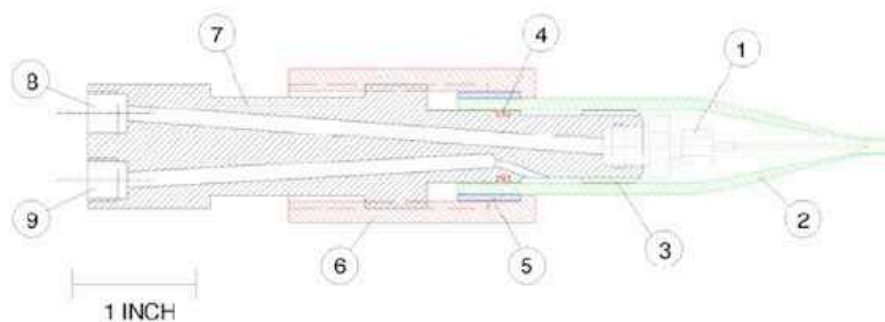

## LIQUID DROPLET BEAM SOURCE WITH COAXIAL GAS FLOW

1 - CAPILLARY NOZZLE, MOUNT, AND FILTER
2 - SHROUD (FUSED SILICA)
3 - VANES TO SUPPORT SHROUD AND STRAIGHTEN GAS FLOW
4 - O-RING SEAL
5 - SHROUD CLAMP RING (GLUED TO SHROUD)
6 - SHROUD CLAMP AND POSITION ADJUSTMENT NUT
7 - MAIN BODY (PEEK)
8 - ANALYTE INLET
9 - SHROUD GAS INLET

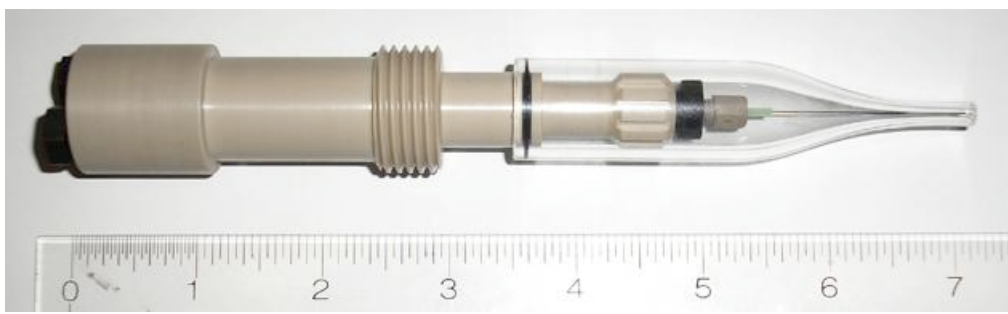

Figure 11: ASU Droplet Beam Source with Coaxial Gas Flow.

The analyte liquid is forced (by a syringe pump or pressurizing gas) out through this capillary nozzle to form the protein-containing droplet beam (our standard way of generating a Rayleigh droplet beam). New to this variant, is the coaxial shroud gas that enters through the lower connection at the end of the main body and is directed into a annular plenum near the front of the body, whence it passes through a set of narrow linear channels (for flow straightening and uniformity) into a fused silica shroud. This shroud has a 15 mm ID at the rear end and reduces to a much smaller (1-3 mm) ID at the front exit. It can be translated back and forth along the axis of the main body to adjust the position of the shroud exit relative to the exit of the capillary nozzle. Any desired shroud gas can be used and the flow velocity is set by choice of inlet pressure. Helium is of particular interest, since (1) it is completely inert; (2) it has very low mass and therefore high diffusivity and thermal conductivity; (3) helium cools extremely effectively during a supersonic expansion into vacuum, and may provide significant cooling of the protein/droplets as they are injected into a vacuum chamber (not shown) downstream of the shroud exit. This shrouded capillary of Figure 11 is being tested to determine the utility of the coaxial flow as a means of drying (evaporatively shrinking) water droplets while still maintaining a linear beam. It is worth noting that gas flow is accompanied by a pressure decrease (Venturi effect). As the coaxial gas flow passes the capillary nozzle at large speeds, it will therefore help to draw the analyte fluid from the nozzle (as in the Gañán-Calvo source [52, 53], see below).

We are not aware of coaxial gas flows having ever been employed for the purpose of drying a Rayleigh droplet beam while maintaining its linear form. There are related prior applications, however. Yang, et al. [54] employed a converging flow of gas to promote gentle evaporation in electrospray ionization of proteins, and suggested that such gas flow improves desolvation by generating smaller droplets which promotes retention of folded protein structures. Their capillaries were much larger (100μm ID and 160μm OD) than those of interest here. Also, they apparently investigated only very high speed flows, at which fragmentation of the liquid is the likely outcome. Electrospray, of course, neither needs nor attempts to produce a single-file, unidirectional, monodisperse droplet beam (although under certain conditions it can produce mono-disperse droplets which are unidirectional close to the nozzle, see below). A coaxial shroud flow of $CO_2$ around an electrospray nozzle is often used to suppress corona discharge from an electrospray capillary. The work by Chen, et al. [55] is typical. Those authors infer a monodisperse droplet distribution from their source (arguing that the residue of droplet solutes after evaporation is monodisperse, and therefore the parent droplets must also be). Again, however, the droplet beam was certainly not single-file or unidirectional and no attempt was made to influence evaporation by adjusting the shroud flow parameters. Supersonic flows of liquid jets, both with and without coaxial gas flows, have been investigated thoroughly as a means of atomizing liquids [56]. Experimental work by Mates et al [57] and Issac, et al [58] are typical. Supersonic atomization nozzles are generally much larger (5-10 mm ID) than those of interest here, and the flow velocities very much higher. The intent of any atomizer is clearly to avoid a unidirectional beam. Many commercial atomizers employ a shroud gas of some sort. (e.g. "MCS Accu Mist" ultrasonic nozzle of Sono-Tek Corp [59]).

## 7. Protein Conformation in Rayleigh droplet beams

It is important to establish that the passage of proteins through a Rayleigh droplet beam does not damage them. The shear stress and pressure acting on the liquid during the ejection process is of concern and could cause pressure-induced denaturing of fragile biological molecules. This is also of concern when inkjet technology is used to generate microarrays of droplets doped with biomolecules. Shear stress during printing of a DNA microarray with a bubble jet printer (thermally generated pressure pulse) has been shown to not cause any damage [60]. The loss in activity of an enzyme, peroxidase, has been measured by printing an aqueous solution using a piezoceramic actuated inkjet printer at different compression ratios [8]. Significant damage to peroxidase has been found even when printing at low compression rates. The damage could be reduced by the addition of trehalose/glucose. Our model system for damage studies was photosystem I (PSI), a light absorbing transmembrane protein complex involved in photosynthesis. Optical absorption spectra from PSI both before and after its passage through a 4μm borosilicate glass nozzle in a Rayleigh droplet beam at two different pressures have been recorded. Both measurements are shown in Figure 12. No damage is evident, since the two spectra are virtually identical. However this does not prove that there were no conformational changes during the flight of the droplets, since the measurement was done on the collected liquid from the Rayleigh jet, i.e. the proteins could have undergone conformational changes during flight. However this is very unlikely, as this complex system cannot self-assemble when it has been disrupted. Further studies, which can probe structural changes during flight are necessary (e.g. spectroscopy during flight, see [61]). Ultimately the serial diffraction experiment itself will be the best indicator.

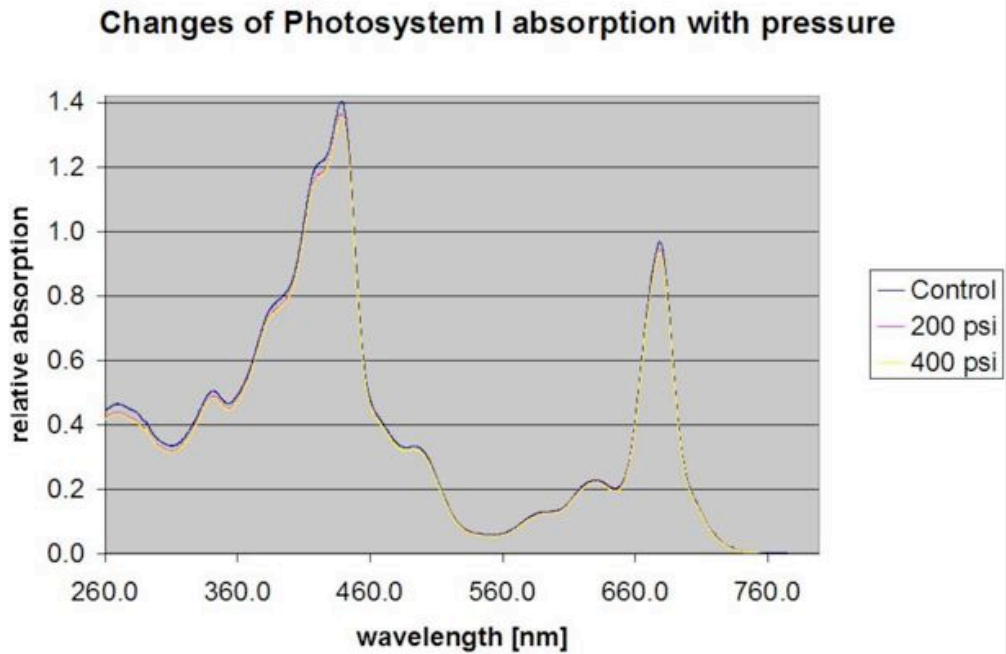

Figure 12: Optical absorption spectra from photosystem I protein complex before (control) and after passing through a 4μm borosilicate glass nozzle in a Rayleigh droplet beam at two different pressures.

## 8. X-ray experiments

We have used direct injection of Rayleigh droplets beams into vacuum as a first test for a serial diffraction chamber, which uses soft x-rays as the diffracting radiation. A schematic of the chamber is shown in Figure 13. The x-rays are focused and monochromatized by a zone plate segment [62]. The size of the x-ray beam at the interaction area is 10 μm. The alignment laser will also be focused to this size, but it has not been installed yet. Figure 14 shows a diffraction pattern obtained with this setup from an untriggered Rayleigh droplet beam doped with 100nm gold balls. The x-ray energy was 530 eV, i.e. the wavelength was 2.34 nm, and the nozzle diameter was 4 μm with a backpressure of 200 psi (liquid flow rate 50 microliter/min). The gold ball concentration in solution was $4.5 \times 10^{10}$ particles/ml, which results in about 12 gold balls in the beam at any time (assuming one 8μm droplet in the 10μm-sized X-ray beam at any time). An Airy's disk pattern of the predicted size from the gold balls can be clearly seen (similar to Figure 5, but with different scale due to the different ball size and X-ray wavelength). The Airy's disk generated by the much bigger water droplets is not visible; it is undersampled because of the experimental geometry and primarily obscured behind the beam stop. Diffraction patterns from solutions containing the membrane protein photosystem I in detergent micelles have also been obtained (from unaligned molecules similar to SAX measurements [62]). However, in this experiment the protein concentration was too low and problems with nozzle clogging due to the detergent buffer used will have to be solved before further measurements with this protein can be conducted.

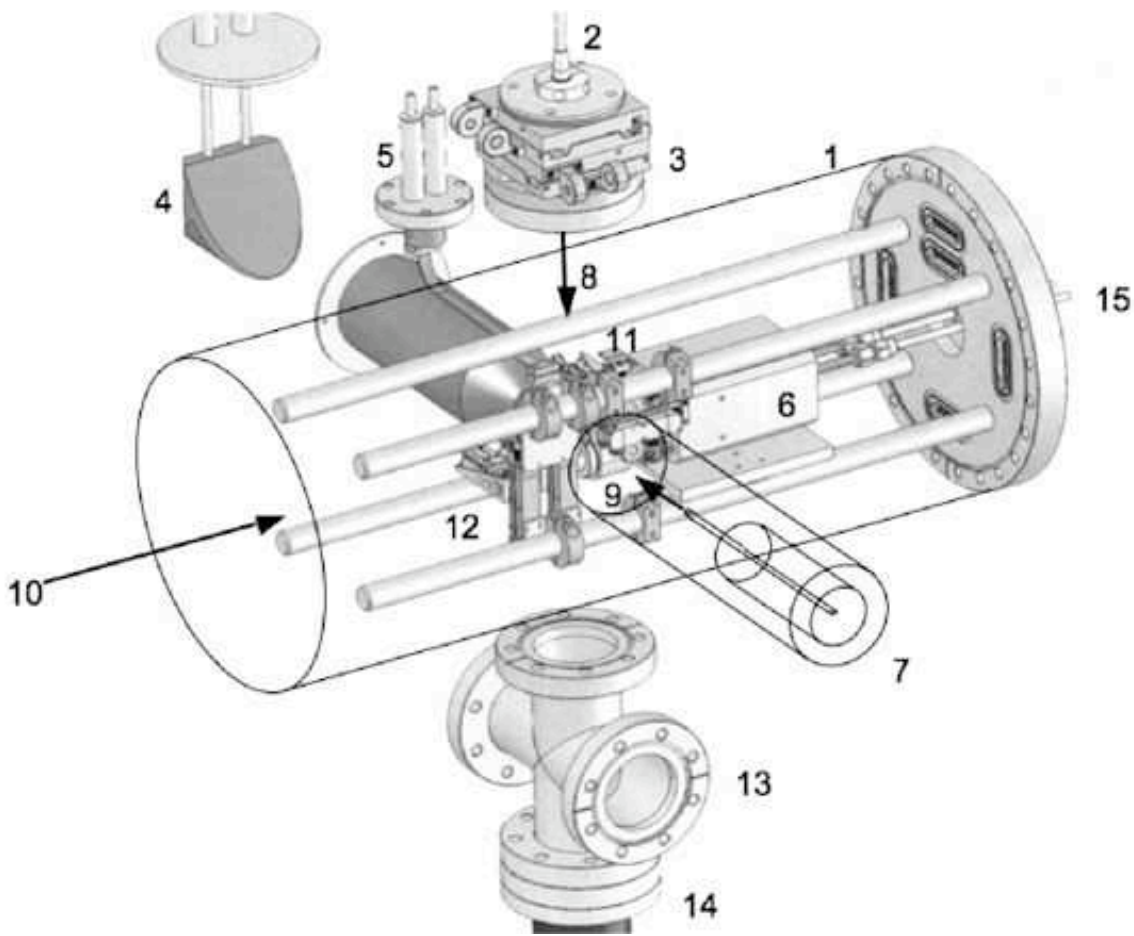

Figure 13: General arrangement of apparatus for Serial Crystallography.
(1) Vacuum Chamber outline, (2) Laser for molecular alignment, (3) Positioning device for alignment of laser beam, laser window and quarter wave plate polarizer (4) $LN_2$ cooled beam dump for droplet beam helps keep the vacuum $<10^{-4}$ Torr, (5) $LN_2$ feedthrough for cryoshield. (6) CCD camera, (7) Protein injection device, (8) Laser beam for molecular alignment, (9) protein beam, (10) X-ray beam, (11) In vacuum lens with positioning motors to focus alignment laser onto protein beam, (12) Motors for alignment of x-ray beam defining apertures, (13) Observation port for visible alignment laser, (14) Water-cooled infra-red laser beam dump, (15) ) Electrical and fluid feedthroughs for CCD and motors.

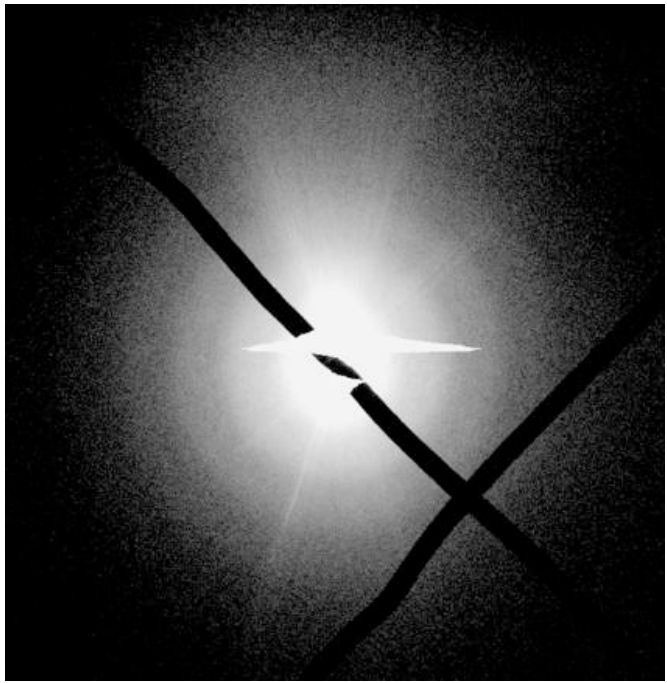

Figure 14: Soft X-ray (530eV, 2.34nm) diffraction pattern from untriggered Rayleigh droplet beam containing 100nm gold balls (possibly within ice, producing radial streaks). A weak Airy's disk pattern originating from the gold balls is visible. Some detector saturation and blooming is visible as horizontal bright streak. The black lines are beam stops for the central beam. The gold ball concentration in the suspension was $4.5 \times 10^{10}$ particles/ml; the nozzle diameter was 4 μm. (the spatial frequency at the edge of the detector is 0.071 $nm^{-1}$ which corresponds to a resolution of 14nm)

### 9. Electrospray

Proteins are routinely injected into vacuum via electrospray ionization (ESI) [50] (see [63] for a recent review), which is a "soft" ionization technique (one that largely avoids fragmentation of the protein during ionization) and is therefore ideally suited for mass spectrometry of large, complex molecules. In ESI, the analyte solution is pumped through a capillary at very low flow rate (0.1 – 10 μl/min). A high voltage (positive or negative) is applied to the capillary. The electric-field gradient at the tip produces a charge separation at the surface of the liquid. As a result, the liquid protrudes from the tip in a "Taylor cone" [64]. Upon reaching the Rayleigh limit [65, 66] (when the Coulomb repulsion of the surface charge is equal to the surface tension of the solution), droplets with an excess charge detach from its tip. In ESI-Mass spectroscopy (ESI-MS), these droplets move through atmospheric pressure gas and generate vapor phase analyte ions. The two mechanisms proposed for ion formation from charged droplets are coulomb fission [67] or alternatively ion evaporation [68]. The continuous flow of charge from the HV power supply through the metal contact to the liquid occurs via an electrochemical reaction at the contact, i.e. oxidation in positive ion mode and reduction in negative ion mode. The analyte ions are transferred from the atmospheric pressure region into the vacuum of the Mass Spectrometer via differential pumping stages (see chapter 3 of [69] ). Formation of cluster ions by condensation of polar neutrals (water and solvent) on the analyte ions during the free-jet expansion into vacuum is prevented by counter current flow of nitrogen gas before the ions enter the vacuum. The gas flow excludes neutrals and water vapor from entering the vacuum system, whereas the ions are attracted by an electric field. The gas flow also helps in ion desolvation.

Initial ESI-MS experiments with proteins in the late 1980s and early 1990s were carried out in high proportions of acidified organic solvents to aid the spray formation and desolvation process. Consequently, the mass spectra recorded were of proteins already denatured in solution. The advent of nano-electrospray enabled ESI-MS from aqueous solutions with smaller droplet sizes and lower charge states [70]. Since then it has been shown that proteins and even macromolecular complexes like the ribosome [71] (for a review see [72]) as well as whole viruses [73, 74] can be transferred into the gas phase via electrospray without fragmentation. A number of protein systems have been investigated using ESI-MS [75-78], and it has been suggested that changes in protein tertiary structures in solution can be monitored using ESI-MS. Depending on experimental conditions, multi-subunit proteins have been shown to stay partially intact and could be selectively dissociated into their subunits [79].

A potential disadvantage of the electrospray approach is that, as the liquid evaporates in these highly charged droplets, the electric field around the particle intensifies. This field may interfere with the protein conformation, since the 3-D structure of proteins is stabilized by electrostatic interactions, which are disrupted by charging the protein. Therefore high charge leads to unfolding and loss of biological activity, well before even more dramatic effects, such as fragmentation, occur. The charge states and the width of the distribution of charge states are commonly used as indications of the degree of unfolding of proteins [80, 81]

To show that proteins are not irreversibly damaged after transferring them as ions into the gas phase, the activity of insulin has been tested after passage through a mass spectrometer [82]. Direct measurements of gas phase protein conformations during flight include infrared spectroscopy [61] and ion mobility measurements [83, 84]. Infrared spectroscopy has shown that the spectra obtained from ESI gas phase proteins (bovine cytochrome c) in low charge states are similar to those in solution, suggesting a similar conformational distribution [61], although changes in secondary structure could not be ruled out. Ruotolo et al . [84] have shown that a small fraction of very stable proteins in their lower ionization states may maintain their 3-D structure, however even in these experiments, more than 80% of the protein was denatured.

Under certain conditions of pressure and voltage, electrospray can produce monodispersed droplets [55]. Usually electrospray is used with little or no additional (other than atmospheric) pressure on the liquid (for a review of electrospray modes see [85]). The droplets generated by electrospray are in general much smaller than the diameter of the nozzle. The droplet size in electrospray depends on the size of the zone at the tip of the Taylor cone where the droplets are emitted. This emission radius in turn is proportional to the 2/3 power of the flow rate dV/dt [86]. Therefore the droplets produced with conventional electrospray sources with nozzle sizes of 20-100 micrometer diameter and flow rates of 1-10ul/min [87] have an initial diameter of 1 - 2 µm whereas droplets produced by nano-electrospray with 1-10µm diameter nozzles and low flow rates of 20 - 40nl/min [88] can be less than 200 nm in diameter. We have done experiments with nano-electrospray nozzles (1 – 8µm diameter) at high liquid pressure (20-70 psi) where larger flow rate produces larger (micrometer sized) droplets, which can be observed with an optical microscope. For initial X-ray experiments the goal was to create 1 – 2µm diameter droplets by direct injection into vacuum. A Rayleigh droplet beam from a 0.5 – 1µm nozzle at 400 psi would deliver the same droplet size but clogging problems with Rayleigh beams are much more serious. A special vacuum flange with the nozzle situated between two windows has been built to allow observation of the electrospray source in vacuum with an optical microscope. A radiative heater consisting of a tungsten wire loop was mounted close to the nozzle to prevent ice formation [89].

Droplets generated by the electrospray source have been imaged with exposure times of 100 ns produced by pulsing a laser diode as described above (see [90] for a similar image of electrospray droplets from a larger nozzle). Figure 15 shows electrospray droplets originating from a 1µm diameter capillary where the liquid (water) was pressurized at 15 psi. This pressure is too low for the spontaneous start of a laminar (Rayleigh) jet, but it is higher than usually used in ES. The flow rate in our experiment was about 500 nl/min. The droplets produced at this high flow rate from a 1µm nozzle are about 1 µm diameter as seen in Figure 15. Initially the droplets appear monodispersed and unidirectional, and then they slow down due to air drag and Coulomb forces, which disperse the droplets to a mist. In vacuum the droplets keep their unidirectionality for longer as shown in Figure 16.

A comparison of electrospray droplets and a Rayleigh droplet beam from the same 8μm diameter nozzle is shown in Figure 17. It is obvious that small droplets can be produced in electrospray mode (from a large nozzle) without the clogging problems associated with the generation of an equally sized Rayleigh droplet beam. Figure 18 shows a comparison of electrospray droplets with different liquid pressure (flow rate). Since the Taylor cone emission radius, which determines the droplets size, is proportional to the 2/3 power of the flow rate dV/dt, the droplet size can be varied by varying the flow rate.

ESI experiments with photosystem I protein in detergent buffer were not successful, since no stable spray could be established. This is probably due to the differences in surface tension and electrolyte concentration. Therefore Rayleigh droplet beams where used instead for our first X-ray experiments.

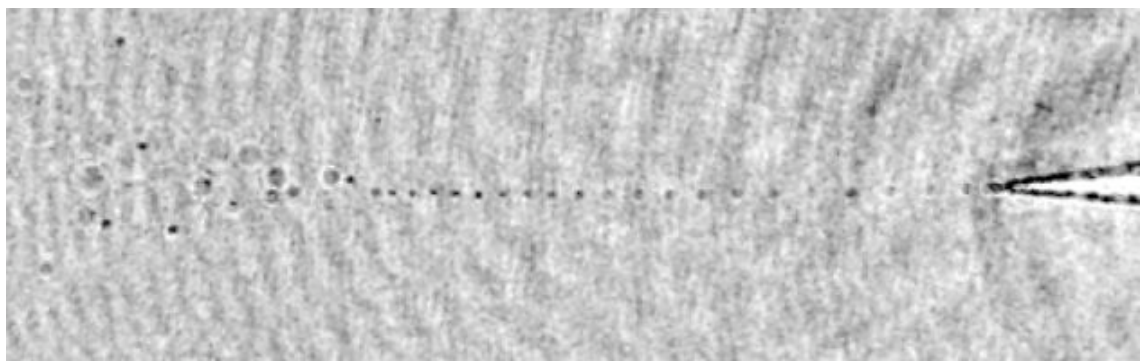

Figure 15: Single shot exposure (100ns exposure time) of electrospray droplets from 1μm diameter nozzle into stagnant air. Dispersion starts about 200 μm downstream of nozzle due to deceleration in air and coulomb repulsion. Droplets appear initially monodispersed, unidirectional and equidistant. Pressure: 15 psi; High voltage: 1.3 kV.

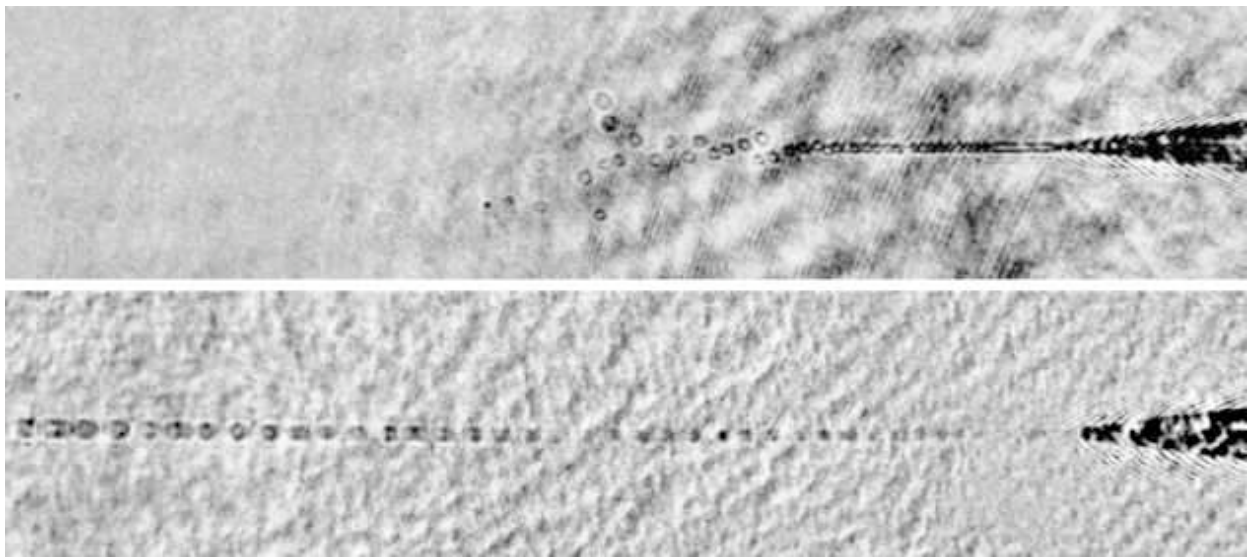

Figure 16:
Top: electrospray of water into stagnant air. Dispersion of the jet is due to slowdown caused by air drag and coulomb forces between drops. Drops in the dispersion region are much slower than at the nozzle. This can be visualized by using a long time exposure time (~700 ns). Then the droplets close to the nozzle were washed out to an apparent jet, but the droplets in the breakup region could still be resolved. Nozzle diameter 1 μm, 2 kV, 70 psi. Single shot exposure 100 ns with laser diode.
Bottom: ES in vacuum, p = $10^{-3}$ Torr, same nozzle, 2 kV, 70 psi. To start the electrospray, the nozzle was heated in vacuum with a tungsten filament close to the nozzle (to thaw the ice generated by evaporative cooling). The HV has to be turned off during pumpdown to prevent discharges in the low vacuum corona discharge region. The droplets stay unidirectional at least up to the extraction electrode (about 2 mm). The laser diode light (50 mW) was focused into the vacuum chamber just below the droplet stream by a lens.

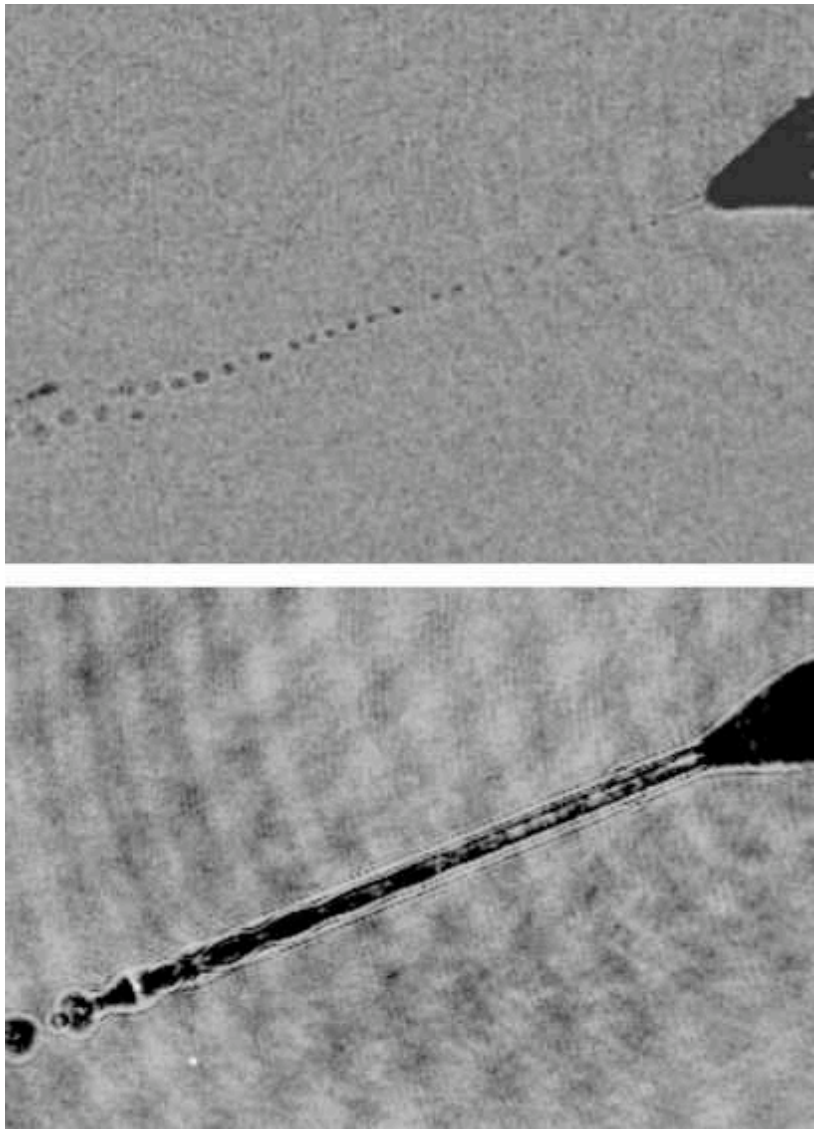

Figure 17:

Top: Electrospray of water from 8μm diameter nozzle at 2.6 kV and 40 psi water pressure (water pressure was too low to start a jet without voltage). Droplets are in focus at the nozzle and out of focus further downstream.

Bottom: Triggered Rayleigh Droplet Beam from same 8μm nozzle at 100 psi and 600kHz trigger frequency (no voltage applied). Diameter of jet: 8 μm, Droplet diameter ~16 μm. Both images are single shot with 100ns exposure time.

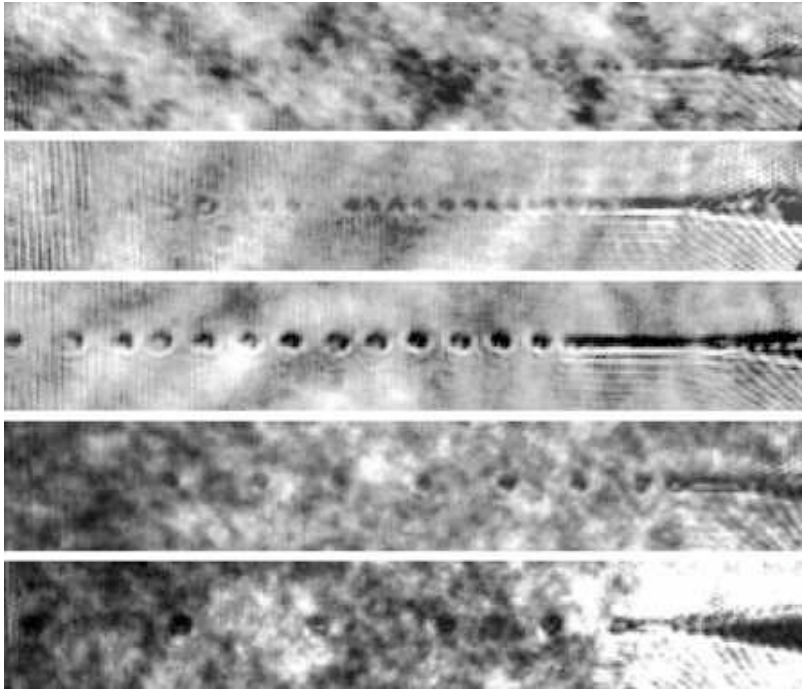

Figure 18: Electrospray from 1μm diameter nozzle at 1 kV in air, pressure from top to bottom: 10, 20, 40, 50, 70 psi. Droplet spacing increases and droplets become larger at higher flow rates (exposure time 100 ns, background is due to coherent speckle pattern of laser illumination).

## 10. Nebulizer

Another way to avoid clogging problems with small Rayleigh nozzles while still obtaining small droplets is the use of a nebulizer [91]. These are straight-bore capillary tubes within a rapidly converging concentric glass shroud. The capillary extends either right up to or to within ~0.5 mm of the end of the concentric shroud, with only a very small annular gap (~20 μm) between the two. The fluid is drawn out by suction (aspiration) alone at liquid flow rates of 10-1000 micro-liter/s.

The smallest droplets generated by these nebulizers can be of the order of 1μm diameter, but the droplets are not unidirectional or mono-dispersed. However, since it produces small droplets with relatively large nozzle sizes, it does avoid the clogging problem. If the shroud gas is helium, these droplets could be effectively cooled by expansion through a skimmer into vacuum. Figure 19 shows an image of the droplet breakup from a commercial nebulizer [92] with a capillary diameter of 150 μm taken with 100ns exposure time.

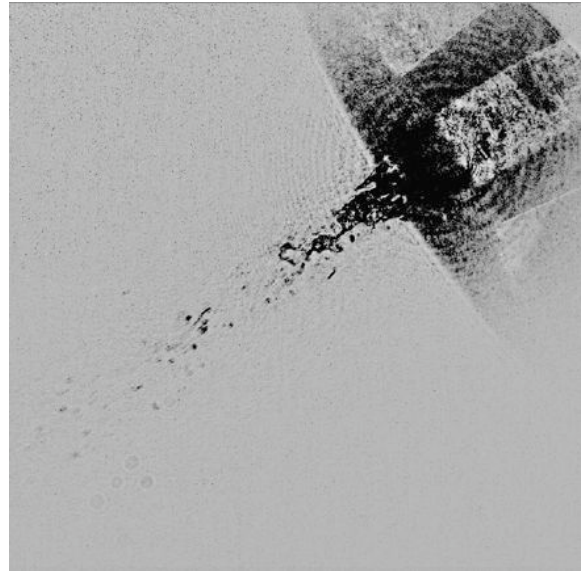

Figure 19: Fast exposure (100 ns) of droplet breakup with a nebulizer. Droplet diameters between 20 μm and 4 μm. Dimensions of the nebulizer: ID and OD of capillary (containing the liquid) are 150 μm and 180 μm respectively. ID of coaxial gas shroud: 230 μm. Gas pressure: 80 psi, liquid pressure: 170 psi through 0.5μm filter.

## 11. Nozzle Cleaning and Filtering

All techniques of liquid jet generation based on nozzles run afoul of clogging at small nozzle diameters. This is particularly true of techniques that require large volumes of liquid to be driven through a converging nozzle by hydrostatic pressure. In such applications a nozzle diameter of 10 μm or even larger is often accepted as the minimum working size. Although still smaller converging nozzles are often employed as electrospray emitters, the driving force in that case is an electrostatic force ("pulling" the liquid out of the nozzle from the front) rather than hydrostatic pressure ("pushing" the liquid out from the rear). Indeed, we have found that nozzles will often run in electrospray mode even after they have clogged in the pressure-driven Rayleigh jet mode. Electrospray volumes are also miniscule, often in the picoliter range for "offline" applications, so that even a short working time before clogging may suffice to carry out an experimental measurement.

For the protein structure determination scheme outlined above, the diffracting beam clearly should not suffer appreciable attenuation in passing through the droplet. Hence the penetration depth of the probe beam through the droplet liquid places an upper limit on droplet size: For x-ray and electron transmission through water at the energies of interest, the maximum droplet diameter is roughly 1 μm and 100 nm, respectively. Calculations of nanoscale liquid jet behavior [93] show Rayleigh break-up of a liquid jet to persist down to nozzle diameters of only a few nanometers, indicating that there is no physical barrier to forming Rayleigh droplet beams of very small droplets. There is a technical barrier to droplet size, however, imposed by nozzle clogging. Small aerosol particulates are ubiquitously abundant in the atmosphere, in concentrations that increase dramatically with decreasing particle diameter below 1 μm. Average US atmospheric concentrations for particulates of under 2.5μm diameter ($PM_{2.5}$) and under 10 μm ($PM_{10}$) [94] are about 13 μg/m$^2$ and 23 μg/m$^2$, respectively. $PM_{2.5}$ particulates are readily suspended in air, carried and traveling like gaseous species. The numbers above give upper estimates on ambient particulate concentrations of about $10^{-2}$, 10, and $10^4$ particles/cm$^3$ for particles of diameter 10, 1, and 0.1 μm, respectively (computed for monodisperse distributions of aerosol matter having density 1.5 g/cm$^3$). These limits suggest that clogging of 10μm nozzles by ambient particulates should be easily avoided whereas the cleaning and filtering needed to avoid clogging of 0.1μm nozzles will present significant technical challenges. It is worth noting, however, that current integrated circuit linewidths are considerable less than that 0.1 μm. Hence the microelectronics industry routinely prevents contamination by particles of this size.

Given ambient concentrations for 1μm diameter particulates of a few particles/cm$^3$ or less, it would seem entirely feasible to develop cleaning and filtering techniques that would allow reliable generation of liquid jets from 1μm nozzles, which would be a factor of ten smaller than the usual working limit. Accordingly, we have devoted considerable effort to developing and testing such techniques. That work is described in detail elsewhere [95]. Stainless steel nozzle tubes were investigated, as were borosilicate glass and fused silica nozzles. PEEK fittings were exclusively employed to mount and connect the glass and silica nozzles, including both commercial fittings [45] and custom-made parts. The operative assumption was that nozzles, once cleaned and protected by a filter on the upstream side, could be stored and then eventually used with no further risk of contamination. This presupposes that (1) all particulates smaller than the nozzle diameter can be removed from the nozzle tube by an initial cleaning and flushing, (2) the filter element can be attached to the nozzle without introducing new particulates into the nozzle tube, and (3) the filter does not itself pass or shed contaminant particles. In addition, it is necessary when operating with a biological solution that (4) the working fluid (which may be a complex chemical solution, possibly containing detergents and other chemically active species) does not clean contaminants from the nozzle wall and (5) that the proteins within the solution do not clump or coagulate in sizes large enough to clog the nozzle.

Equipment and techniques were developed to flush nozzles from both the distal end and the tip end. Numerous cleaning solutions were investigated, both with and without ultrasonic excitation. HPLC water and research grade solvents were used in this cleaning and all liquids were filtered multiple times through commercial 0.5μm sintered PEEK filters. Plasma cleaning of nozzles was explored. Fire polishing techniques were developed to remove sharp edges from the distal ends of nozzle tubes and from the front of the fiber optic capillaries used for back-flushing. Cleaning and flushing of capillaries prior to "pulling" of the nozzle was tried. Sol-gel techniques were employed to grow silicate frit filters within the nozzle tube just upstream of the nozzle. Many of these techniques did bring about improvements in performance, but even in their aggregate they did not provide the desired reliable operation desired in the sub-4 μm range of nozzles. Nozzles of 4 μm would "turn on" fairly reliably a first time, even with only moderately careful preparation. Attempts to re-start such nozzles after the supply reservoir had been emptied and re-filled, however, were successful only about 10% of the time. Nozzles of 8μm diameter, in contrast, would run indefinitely without clogging. It is apparent that even with considerable care in cleaning and filtering - albeit in a standard laboratory setting and not under clean room conditions - dependable operation of nozzles below 4μm diameter (despite the use of much smaller filters) remains extremely challenging. Many-particle interactions may be responsible.

## 12. Aerojet focused source

Given the technical challenges of cleaning and filtering micrometer-sized nozzles, it is worth investigating alternative approaches in which the converging end of a solid nozzle is replaced by a converging aerodynamic flow. One promising approach is that described by Gañán-Calvo [52, 53, 96], which uses fluid dynamical means to generate a liquid cone and small droplets from a large capillary. The liquid is injected through a capillary needle inside a sealed driving chamber. A hole in the driving chamber is located just opposite the needle's end. Gas, in this case helium, is forced through the hole by the extra pressure existing in the chamber. Due to suction forces a meniscus develops (similar to a Taylor cone) from whose vertex a very thin jet is issued. The jet size can be adjusted by adjusting the gas pressure and liquid pressure. This technique avoids clogging problems (like ESI) while producing a (presumably uncharged) collimated beam of droplets much smaller than the nozzle diameter. We are currently investigating these sources for their possibilities as protein injection sources. Figure 20 shows a close-up view of our prototype source in operation. The water meniscus formed at the end of the glass capillary by the gas flow is clearly visible and a jet is ejected, which then breaks up into droplets with diameter about 10 times smaller than the capillary ID (see Figure 21). The droplet breakup in these jets can be triggered by an external piezo transducer, by analogy with the Rayleigh droplet sources, which leads to monodispersed droplets as shown in Figure 22. A current measurement between the source assembly and an electrode in the liquid shows a streaming current about a hundred times smaller than measured with Rayleigh jets, i.e. the droplets produced are almost uncharged. Preliminary experiments with detergent buffer as used for membrane proteins showed the same reliability of the droplet formation process as with pure water. By adjusting the experimental conditions (aperture sizes, pressures) we have recently suceeded in producing collimated droplet jets of micrometer or submicrometer diameter without the clogging problems associated with the production of Rayleigh jets.

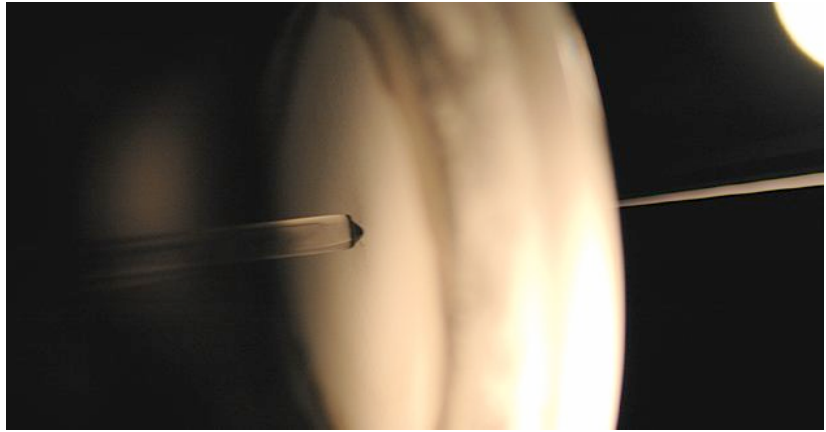

Figure 20: Closeup view of our prototype aerojet focused source. A glass capillary (500μm ID) extends from the left and terminates close to a 500μm ID aperture. Water in the glass capillary is pressurized and helium is forced through the aperture by the extra pressure existing inside the chamber to the left. Due to suction forces a meniscus develops (visible at the end of the capillary) and a fine jet exits to the right.

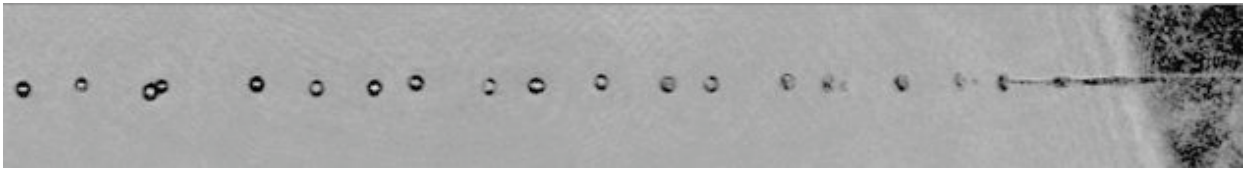

Figure 21: Fast exposure (100 ns) of water droplet jet ejected from aerojet focused source. The source is on the right. The dimensions of the apertures are smaller than in Figure 20: The capillary supplying the water has an ID of 50 μm. The diameter of the jet before breakup is about 5 μm, i.e. a reduction by a factor of 10. The gasflow aperture has a diameter of 100 μm. The droplet diameter is about 10 μm. Water pressure 20 psi, gas pressure 6 psi (measured at the gas supply bottles, the water pressure at the nozzle is reduced due to the use of a 0.5μm filter in the water line). By reducing the water pressure, i.e. the flow rate, the droplet diameter could be reduced to about 5 μm.

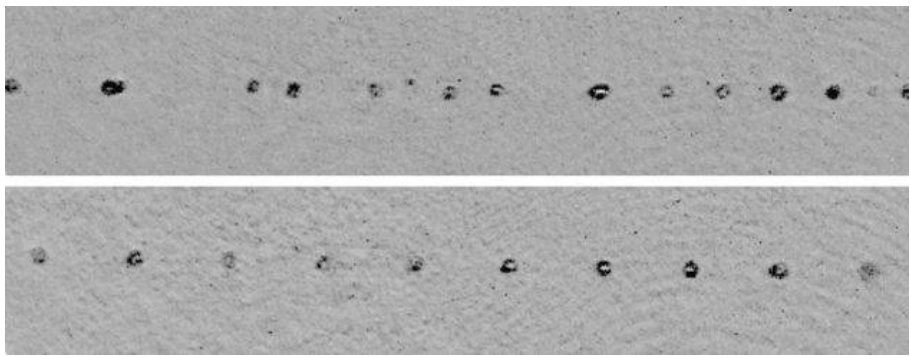

Figure 22: Untriggered (top) versus triggered (bottom) droplet breakup from an aerojet focused source. The source is on the right. Exposure time: 200 ns, water pressure: 25 psi, gas pressure: 5 psi, piezo trigger frequency: 405 kHz, droplet diameter 10 μm. Source dimensions are the same as in Figure 21.

## 13. Acknowledgements
This work was supported by NSF award IDBR 0555845 and ARO award DAAD190010500.
## 14. References

[1] Spence J C H and Doak R B 2004 *Single molecule diffraction* Physical Review Letters **92** 198102
[2] Wu J S, Leinenweber K, Spence J C H and O'Keeffe M 2006 *Ab initio phasing of x-ray powder diffraction patterns by charge flipping* Nature Materials **5** 647-52
[3] Kurkal V, Daniel R M, Finney J L, Tehei M, Dunn R V and Smith J C 2005 *Enzyme activity and flexibility at very low hydration* Biophysical Journal **89** 1282-7
[4] Dunn R V and Daniel R M 2004 *The use of gas-phase substrates to study enzyme catalysis at low hydration* Philosophical Transactions of the Royal Society of London Series B-Biological Sciences **359** 1309-20
[5] Henderson R 2004 *Realizing the potential of electron cryo-microscopy* Quarterly Reviews of Biophysics **37** 3-13
[6] Bartell L S and Huang J F 1994 *Supercooling of water below the anomalous range near 226 k* Journal of Physical Chemistry **98** 7455-7
[7] Howard E I and Cachau R E 2002 *Ink-jet printer heads for ultra-small-drop protein crystallography* Biotechniques **33** 1302
[8] Nishioka G M, Markey A A and Holloway C K 2004 *Protein damage in drop-on-demand printers* Journal of the American Chemical Society **126** 16320-1
[9] Arakawa E T, Tuminello P S, Khare B N and Milham M E 2001 *Optical properties of ovalbumin in 0.130-2.50 mu m spectral region* Biopolymers **62** 122-8
[10] Arakawa E T, Tuminello P S, Khare B N and Milham M E 1997 *Optical properties of horseradish peroxidase from 0.13 to 2.5 mu m* Biospectroscopy **3** 73-80
[11] Simonson T 2003 *Electrostatics and dynamics of proteins* Reports on Progress in Physics **66** 737-87
[12] Koch M H J, Dorrington E, Klaring R, Michon A M, Sayers Z, Marquet R and Houssier C 1988 *Electric-field x-ray-scattering measurements on tobacco mosaic-virus* Science **240** 194-6
[13] Bras W, Diakun G P, Diaz J F, Maret G, Kramer H, Bordas J and Medrano F J 1998 *The susceptibility of pure tubulin to high magnetic fields: A magnetic birefringence and x-ray fiber diffraction study* Biophysical Journal **74** 1509-21
[14] Sutter P-2000, see http://www.sutter.com/products/product_sheets/p2000.html.
[15] Hager D B, Dovichi N J, Klassen J and Kebarle P 1994 *Droplet electrospray mass-spectrometry* Analytical Chemistry **66** 3944-9
[16] Hager D B and Dovichi N J 1994 *Behavior of microscopic liquid droplets near a strong electrostatic-field - droplet electrospray* Analytical Chemistry **66** 1593-4
[17] Hemberg O, Hansson B A M, Berglund M and Hertz H M 2000 *Stability of droplet-target laser-plasma soft x-ray sources* Journal of Applied Physics **88** 5421-5
[18] Chudobiak M J 1995 *High-speed, medium voltage pulse-amplifier for diode reverse transient measurements* Review of Scientific Instruments **66** 5352-4
[19] Frohn A and Roth N 2000 *Dynamics of droplets*. (Berlin: Springer)
[20] Middleman S 1998 *An introduction to fluid dynamics*. (New York: Wiley)
[21] Rayleigh L 1878 *On the instability of jets* Proc London Math Soc **10** 4-13
[22] Hanson E 1999 *Recent progress in ink jet technologies ii*. (Springfield, VA Society for Imaging Science and Technology)
[23] Siegbahn H and Siegbahn K 1973 *Esca applied to liquids* J Electron Spectr Rel Phenom **2** 319-25
[24] Keller W, Morgner H and Muller W A 1986 *Probing the outermost layer of a free liquid surface - electron-spectroscopy of formamide under he(2(3)s) impact* Molecular Physics **57** 623-36
26